\documentclass[%
reprint,
showpacs,preprintnumbers,
amsmath,amssymb,
aip,
jap,
]{revtex4-1}

\usepackage{graphicx} 
\usepackage{amsmath}

\usepackage{dcolumn}
\usepackage{bm}

\usepackage{hyperref}
\hypersetup{colorlinks,urlcolor=blue}

\begin{document}


\title{Electron-phonon interaction in an $N$-atomic periodic chain}



\author{Zahidy, M.}
\email[]{zahedi\_mojtaba@physics.sharif.edu}

\affiliation{Department of Physics, Sharif University of Technology, Tehran, Iran}

\author{Ghadirian, F.}
\email[]{fa.ghadirian@yahoo.com}

\author{Namiranian, A.}
\email[]{afshinn@iust.ac.ir}
\affiliation{Department of Physics, Iran University of Science and Technology, Tehran, Iran}


\date{\today}

\begin{abstract}
The study of electron-phonon interaction as a prominent inelastic effect is of great importance. In the present work, we have studied the inelastic effects due to the first order electron-phonon interactions on electronic properties of mono-atomic periodic chain, using the exact diagonalization technique. Hence, only acoustic modes are considered. To avoid the approximate results arising from Keldysh formalism and on-site electron-phonon interaction assumption, we have used the Green's function technique together with Fr\"{o}hlich Hamiltonian for the interaction part. Finally, as an example, we apply our method to the case of $N=6$ which could be considered as a Benzene-like molecule.
\end{abstract}


\maketitle

\section{Introduction}
Collective vibrations of atoms play an important role in different phenomenas such as in conventional superconductors \cite{PhysRev.108.1175}, anharmonic phonon scattering in thermoelectric materials \cite{Delaire2011}, and in colossal magnetoresistance \cite{1473}. The effect of electron-phonon interactions in molecular electronics has been the subject of extensive study \cite{PhysRevB.68.205323,PhysRevB.73.115405,PhysRevLett.94.206804,PhysRevB.73.045314,Benesch2006355,PhysRevB.76.155430,0953-8984-19-10-103201} and its influence on the conductivity is shown. Moreover, phonon-mediated effects are also detected experimentally \cite{PhysRevLett.88.226801,PhysRevB.72.075413,PhysRevLett.92.206102,Park2000}.

Several numerical methods are devised to tackle the problem of electron-phonon interaction, such as exact diagonalization (ED) \cite{MARSIGLIO1993280,Nath2015,PhysRevB.55.14872}, density matrix renormalization group (DMRG) \cite{PhysRevB.88.224512,Tezuka2005708,PhysRevB.89.155434}, quantum Monte-Carlo (QMC) \cite{PhysRevLett.51.296,PhysRevB.39.5051,PhysRevB.89.195139,PhysRevB.89.195139}. A review of the applications of dynamical mean-field theory (DMFT) in strongly correlated systems can be found in \cite{1063-7869-55-4-R01}.

A wide range of theoretical studies concerning electron-phonon interaction in $N$-atomic systems have used localized phonon assumption, known as Einstein phonon \cite{Tezuka2005708,PhysRevB.89.195139,Matsueda20083070,Nath2015}.
This means that an electron on a specific site interacts only with the phonons on the same site.
The Holstein-Hubbard Hamiltonian \cite{HOLSTEIN1959325} is in this class.
This approach has two obvious drawbacks. On one hand the localized phonon approach is inconsistent with the fact that a phonon has a particular momentum and assigning a position to the phonon is impossible and on the other hand it is in contrast with the existence of various phonon modes in such systems. A single nuclei vibrating under the influence of local potential and independent of the others can be modeled as a single mode harmonic oscillator. Hence, this assumption can lead to inaccurate results.


The approach of non-equilibrium Green's function (NEGF) together with Keldysh formalism is the main approach to study non-equilibrium transport properties in molecular devices.
However taking account of the interaction whether it is of electron-electron type or electron-phonon and finding the self-energies is only possible to some order, which could cast errors in the results when interactions are strong \cite{PhysRevB.79.085120}.

In the theoretical approach presented here, to avoid the approximate results arising from adopting the Keldysh formalism and on-site electron-phonon interaction assumption, we have used the Green's function formalism and tight-binding Hamiltonian for the electronic part together with Fr\"{o}hlich Hamiltonian \cite{frohlichpolaron} for electron-phonon interaction.
While ED provides exact results for finite clusters, DMRG and DMFT are exact in infinite dimensions \cite{RevModPhys.68.13}. DMFT also neglects the spatial correlations. Thus it is more suitable for investigating the Holstein-type Hamiltonian \cite{PhysRevB.89.195139}. Hence, we have used the ED technique to find the Green's function and study the transport properties of the model. 

Calculating the current profile of the system shows that each step in the current breaks into a number of smaller steps in the presence of electron-phonon interaction. This happens according to the dimension of the phonons Hilbert space that could be controlled by changing the temperature. The results from the presented example with $N=6$ which could be regarded as a Benzene molecule with some assumptions, are qualitatively similar to what is observed in IET spectroscopy \cite{Song2009,PhysRevLett.88.226801,Park2000}.
    
This paper is organized as follows. In section II a description of electron-phonon interaction and the related Hamiltonian are given. In section III, we explain our proposed method and finally in section IV we apply our method to the case of $N=6$ and present our results.
\section{Theoretical Description}
We consider an $N$-atomic periodic chain. The total Hamiltonian of this electron-phonon coupled system has the following form
\begin{equation}
	H = H_e \otimes I_{ph} + I_e \otimes H_{ph} + H_{e-ph}.
	\label{eq1}
\end{equation}

The first term $H_e$ describes the electronic part of the system in tight-binding formalism
\begin{equation}
	H_e = \sum_{i,\sigma} \epsilon_i c_{i,\sigma}^{\dagger} c_{i,\sigma} - t (c_{i+1,\sigma}^{\dagger} c_{i,\sigma}+h.c.),
	\label{eq2}
\end{equation}
where $t$ is the hopping parameter between adjacent sites, $\epsilon_i$ is the on-site energy of a localized electron and $c_i^{\dagger}$ ($c_i$) is the creation (annihilation) operator of an electron at site $i$ with spin $\sigma$. The spin does not play any role in this problem, so we have dropped it in the following expressions.
The second term $H_{ph}$ is the vibrational mode of the system which is described by
\begin{equation}
	H_{ph} = \sum_{i,\lambda} \hbar \omega_{\lambda,q_i} (b_{\lambda,q_i}^{\dagger} b_{\lambda,q_i}+\frac{1}{2} I_{ph_{\lambda,q_i}} ).
	\label{eq3} 
\end{equation}

In (\ref{eq3}), the operator $b_{q_i}^{\dagger}$  ($b_{q_i}$) creates (destroys) a phonon in mode $q_i$ with energy $\omega_{q_i}$ and index $\lambda$ runs over different phonon branches.
In practice, we set aside the constant term of $H_{ph}$ since it only amounts to a shift in energies Also we drop the index $\lambda$ since we just consider the acoustic branch due to dealing with mono-atomic Bravais lattice.
The last term of (\ref{eq1}) represents the interaction between the vibrational modes and the conduction electrons. The Fr\"{o}hlich Hamiltonian for electron-phonon intraction is given by this expression \cite{frohlichpolaron}
\begin{equation}
	H_{e-ph}=\sum_{q,k} \lambda (q) c_{k+q}^{\dagger} c_k (b_{-q}^{\dagger} + b_q ),
	\label{eq4}
\end{equation}
where $c_{q}^{\dagger}$ and $c_q$ are the fermionic creation and annihilation operators related to the electronic level with momentum $q$, and  $b_{q_i}^{\dagger}$ and $b_{q_i}$ are as mentioned before.
$\lambda$ which is a function of phonon momentum represents the electron-phonon coupling strength and is equal to
\begin{equation}
	\lambda_i (q)=\sqrt{\frac{\hbar}{2mN\omega_{q_i}} },
	\label{eq5}
\end{equation}
with $\hbar$ the reduced Planck constant, $m$ as the mass of each nuclei and $N$ as the number of sites in the system. Note that the bosonic operators in (\ref{eq4}) are acting in momentum space. 

It is clear that to use the ED technique, a restriction on the dimension of the Hilbert space is required. This manifest itself in two ways: a restriction on the number of lattice sites and a restriction on the number of phonons \cite{MARSIGLIO1993280}. One can assume that the interactions are weak enough and consider only one phonon per site or mode \cite{Nath2015,Mondal20113723}. Introducing a cutoff on the dimension of the phonon Hilbert space could result in errors. In \cite{PhysRevE.89.042112}, it has been argued that a system deviate from thermalization with a universal power law $D^{-1/2}$ with the dimension $D$ of the Hilbert space. 
Zhang \textit{et al.} \cite{PhysRevLett.80.2661,PhysRevB.60.14092} first proposed a technique to generate a controlled truncation of the Hilbert space and studied the Holstein model. Similar to the key idea in DMRG, the main idea is to discard the state with low probability.

Here we adopt the idea that the excitation probability of phonons is reduced drastically with increasing the energy, thus we have assumed that each mode, on average, is occupied in accordance with Bose-Einstein distribution function, $N_{BE}$($\omega$), which on one hand, is more realistic for bosonic systems and on the other hand gives us the ability to tune the number of phonons with temperature.

Due to the interaction term in the total Hamiltonian, Keldysh formalism can only be used with some approximation \cite{MolecElectrnoic}. However, in the case of strong coupling, this could result in error \cite{PhysRevB.79.085120}.
In this paper, we study the electronic properties of the described system by the use of Green's function formalism. As we mentioned, in general, it is not possible to solve the problem exactly because of the electron-phonon interaction and semi-infinite chain. Thus we adopt the Cluster Perturbation Theory (CPT) \cite{PhysRevLett.84.522}, and divide the molecular device into three clusters: (i) the central region consisting the $N$-atomic chain with vibrational degrees of freedom, (ii) the left and (iii) the right metallic leads which are connected to the bias voltage. 
For the leads, we consider semi-infinite chain with hopping parameter $t_L$, on-site energies $\epsilon_L$ and chemical potential $\mu_L$.
The self-energies due to connection of the contacts to the central region is given by
\begin{equation}
	\Sigma_{L,R}=\tau_{L,R}^2 g_{lead} (E),
	\label{eq6}
\end{equation}
where $\tau_{L(R)}$ is the hopping parameter between the left (right) contact and the central region and $g_{lead}(E)$ is the surface Green's function of the leads. 
From (\ref{eq1}) the Green's function of the isolated central region is obtained as follow
\begin{equation}
	\tilde{G} = [(E+i\eta)I-\hat{H}]^{-1}.
	\label{eq7}
\end{equation}

This is the Green's function of the electron-phonon interacting system. By taking partial trace over phonon's subspace, the electronic part of the central region's Green's function could be obtained. 
Dyson's equation yields the total Green's function i.e. the central region under the influence of leads and external bias. The total Green's function is given by
\begin{equation}
	G=\tilde{G}+\tilde{G}(\Sigma_L+\Sigma_R)G.
	\label{eq8}
\end{equation}

Once the total Green's function of the central cluster, (\ref{eq8}), and the surface Green's function of the leads are obtained, the electronic properties of the system such as local density of states at site $i$, transmission, and current are then calculated by \cite{MolecElectrnoic,PhysRevLett.57.1761,PhysRevB.67.165326}
\begin{equation}
	LDOS_{i}= -\frac{1}{\pi}  \mathrm{Im} \{G(i,i)\},
	\label{eq9}
\end{equation}
\begin{equation}
	\mathcal{T}=\Gamma_L  G \Gamma_R  G^{\dagger},
	\label{eq10}
\end{equation}
\begin{equation}
	I(V)= \frac{e}{\pi h} \int dE [F_L (E,V)-F_R (E,V)]tr[\mathcal{T}(E,V)].
	\label{eq11}            
\end{equation}

In the above equations $F_{L,R} = [e^{(\varepsilon-\mu_{L,R})/k_BT}+1]^{-1}$ is the Fermi-Dirac distribution function of the left and right lead where $K_B$ is the Boltzmann constant and $T$ is the temperature. And $\Gamma_{L,R}=-2 \mathrm{Im} \{\Sigma_{L,R}\}$ is the imaginary part of self-energy. In principle, it is possible to formulate these equation according to advanced Green's function but in this paper, however, we use the retarded Green's function.
Moreover, the current-current correlations (shot noise) in the limit of $\omega \to 0$ is given by the expression \cite{PhysRevB.88.054301}:
\begin{equation}
\begin{split}
	S & = \frac{e^2}{h} \int dE [F_L(1-F_L)+F_R(1-F_R)]\mathcal{T} \\ 
	&+ (F_L - F_R)^2 tr[\mathcal{T}(1-\mathcal{T})].
\end{split}
	\label{eq-shot}            
\end{equation}

\section{The Method}

As mentioned in the last section, two terms of the Hamiltonian (\ref{eq1}), bare electronic term and the interaction term are represented in different basis. In order to solve the problem and find the total Green's function of the system, we need to express both terms in the same basis. Since the interaction term in momentum representation clearly shows the process of electron-phonon interaction, this representation is chosen. Thus we need to write the tight-binding Hamiltonian in momentum representation. The entries of the eigenstates of the tight-binding Hamiltonian determine the probability amplitude which localized orbitals in corresponding sites are occupied. Since there is no unitary transformation between localized orbital basis and coordinate basis, the following approximation has been made. First notice that the eigenstates of the tight-binding Hamiltonian are in the form of
\begin{equation}
	| \Psi \rangle = \sum_{i=1}^N \alpha_i | \phi_i \rangle,
	\label{eq12}
\end{equation}
in which $| \phi_i \rangle$ is the localized orbital at site $i$ and $\alpha_i$ is probability amplitude of occupation associated with each localized orbital. To each orbital, there corresponds an expansion in coordinate basis in the form of
\begin{equation}
	| \phi_i \rangle = \int dr \langle r | \phi_i \rangle | r \rangle.
	\label{eq13}
\end{equation}
which when used, one can express $ | \Psi \rangle$ in coordinate space as
\begin{equation}
| \Psi \rangle = \sum_{i=1}^N \int dr [\alpha_i \langle r | \phi_i \rangle] | r \rangle.
\label{eq13-2}
\end{equation}

It must be borne in mind that in tight-binding Hamiltonian, it is assumed that the adjacent orbitals have no overlap on each other \cite{PhysRev.94.1498}.

In practice, each orbital can be approximated by keeping as many terms in the above expansion as we demand.
Preserving more terms makes the approximation more realistic and precise but it comes with a price. The computational resource for this task grows rapidly since there are $N$ different sites.
In the first approximation we only take one coefficient and omit the others, thus every orbital represents with a Dirac delta function.

Representing the eigenstates in coordinate basis, one can obtain the unitary transformation to the momentum basis
\begin{equation}
	\hat{U} | \Psi_i \rangle = | \phi_j \rangle, \hspace{3 mm} \hat{U}_{m,n}= \frac{1}{\sqrt{N}} e^{ik_m x_n},
	\label{eq14}
\end{equation}
\begin{equation}
	k_m=\frac{2\pi}{Nd}m, \hspace{3 mm} m=0 ...N-1.
	\label{eq15}
\end{equation}

In (\ref{eq14}) and (\ref{eq15}) the $k_m$’s are allowed momentums of the electrons in the system and $d$ is the inter-atomic distance. 

%
%
The main idea is to represent the eigenvectors in coordinate space and then find the unitary transformation to the momentum space.

Finding $\hat{U}$, now it is possible to write the Hamiltonian in momentum representation
\begin{equation}
	\tilde{H}_e = \hat{U} \hat{H}_e \hat{U}^{\dagger}.
	\label{eq17}
\end{equation}

Note that, adopting the above assumption, the tight-binding Hamiltonian interprets as coordinate representation Hamiltonian. Keeping more terms in the expansion (\ref{eq13-2}) we expect more precise results. 

One important remark about the Green's function to remember is when the Hamiltonian comprises of two non-interacting part. Assume that the given Hamiltonian is
\begin{equation}
	H = H_e \otimes I_{ph} + I_e \otimes H_{ph}.
	\label{eqremark1}	
\end{equation}

The eignestates of $H$ are in the form
\begin{equation}
	|\Psi_{m,n} \rangle = | \psi_m^e \rangle \otimes | \psi_n^{ph} \rangle,
	\label{eqremark2}	
\end{equation}
with eigenenrgies
\begin{equation}
	E_{m,n}= E_m^e + E_n^{ph}.
	\label{eqremark3}	
\end{equation}
Calculating the Green's function of $H$ gives
\begin{equation}
\begin{split}
G(H,E) & = \sum_{m,n} \frac{(| \psi_m^e \rangle \otimes | \psi_n^{ph} \rangle)(\langle \psi_m^e | \otimes \langle \psi_n^{ph} | )}{E-E_{m,n} \pm i \eta} \\
& = \sum_{n}[\sum_{m} \frac{| \psi_m^e \rangle \langle \psi_m^e |}{E-E_m^e-E_n^{ph} \pm i \eta}] \otimes | \psi_n^{ph} \rangle \langle \psi_n^{ph} | \\
& = \sum_{n=1}^{D_{ph}} G^e(E-E_n^{ph}) \otimes | \psi_n^{ph} \rangle \langle \psi_n^{ph} |,
\end{split}
	\label{eqremark4}	
\end{equation}
where $D_{ph}$ is the dimension of phonon Hilbert space.

Taking the partial trace over the phonon space, one should be able to find the electrons Green's function. One can see that in Eq.(\ref{eqremark4}) the electronic part of Green's function experience shifts according to the phonon eigenenergies and then added up together.
Since there is no interaction between electrons and phonons in (\ref{eqremark1}), we expect to see no changes in electronic Green's function and consequently in transport properties.

Assuming the interaction strength is low enough, using perturbation theory one is able to calculate the correction of eigenenergies due to the interaction. The correction is given by
\begin{equation}
	E = E_0^e + \frac{\lambda^2(q_i) n_i}{\pm \hbar \omega_i},
	\label{eq18}	
\end{equation}
in which $E_0^e$ is the eigenenergy of bare electronic system and the ($\pm$) correspond to emission or absorption of a phonon respectively. In the absence of interaction, each eigenstate of the system has a degeneracy according to dimension of the phonon’s Hilbert space. As we show in the next section and according to (\ref{eq18}), the effect of interaction is to break this degeneracy.


\section{Example and Results}

\begin{figure}[!h]
	\centering
	\includegraphics[width=0.5\textwidth]{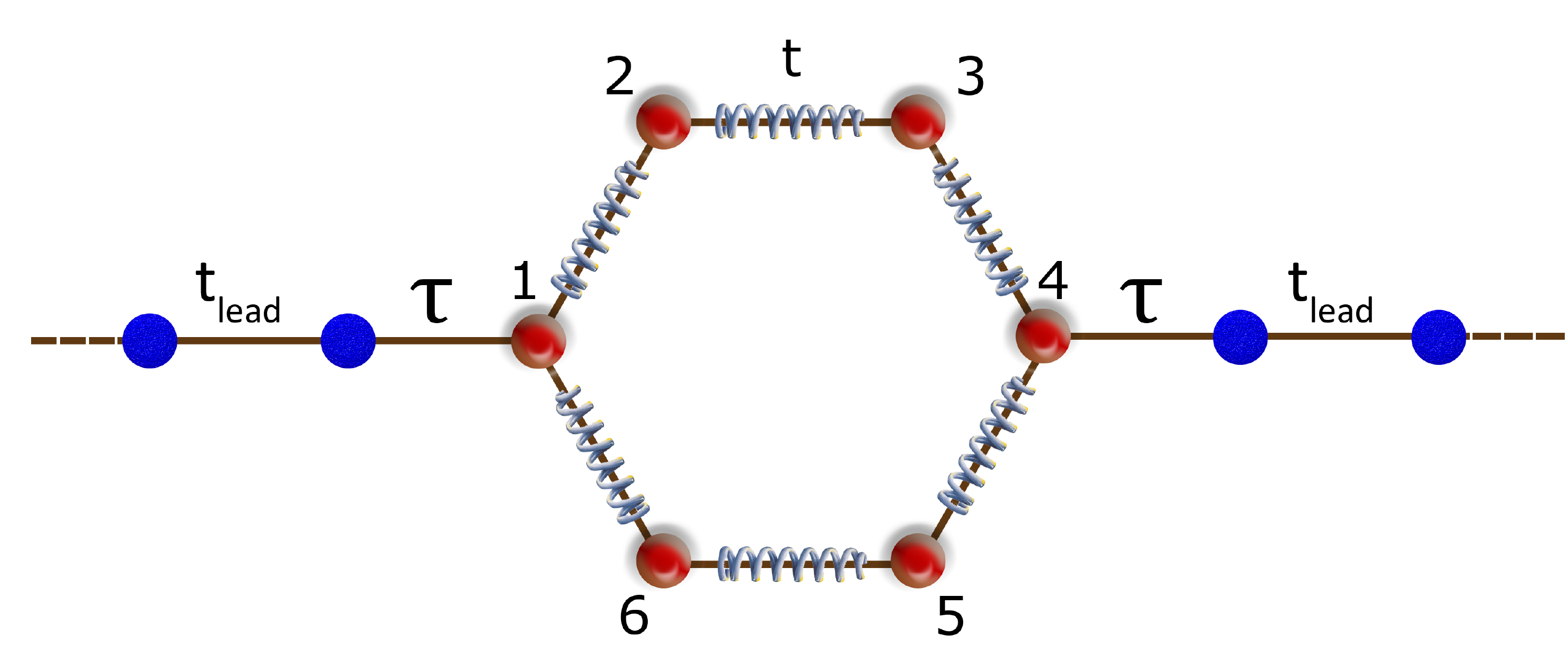}
	\caption{Illustration of the periodic chain (red spheres) with hopping parameter $t$, and leads (blue spheres) with hopping parameter $t_{lead}$ and tunneling coupling constant $\tau$ between leads and central cluster.}
	\label{fig1}
\end{figure}

We apply our method to the case of $N=6$ and investigate the effect of electron-phonon interaction on the transport properties of this periodic chain. An illustration of the system which is connected to semi-infinite metallic chains as leads is given in Fig. (\ref{fig1}).

\begin{figure}[!h]
	\centering
	\includegraphics[width=0.5\textwidth]{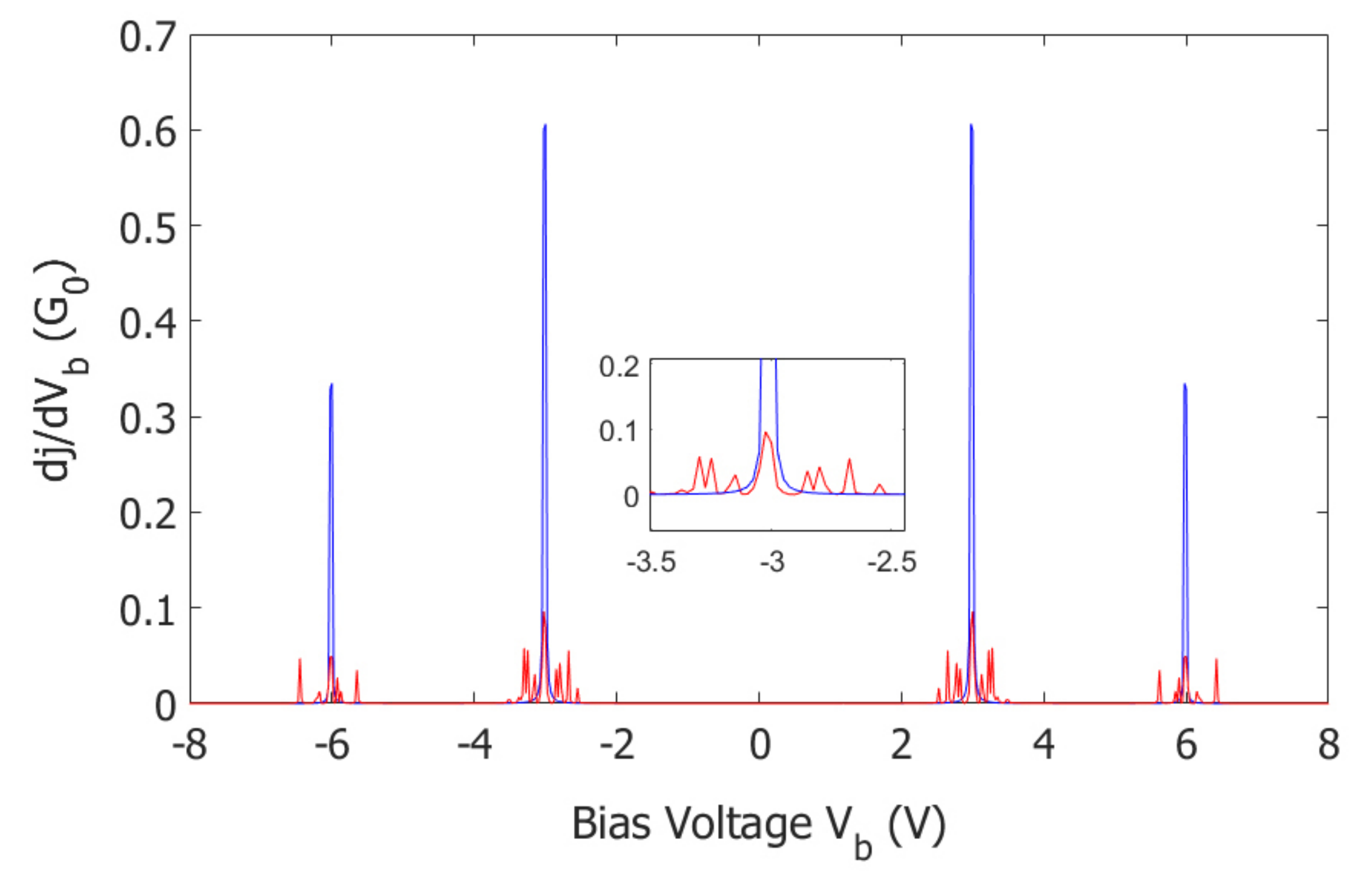}
	\caption{Illustration of conductance in a rigid chain (blue) and in the presence of electron-phonon interaction (red) as a function of bias voltage. The four transmission peaks centered at bias voltages equal to eigenenergies $E=\pm 3 eV$ and $E=\pm 6 eV$ break into a number of smaller peaks due to electron-phonon interaction. The inset shows the conductance at bias voltage $\text{V}_b=-3 eV$.}
	\label{fig2}
\end{figure}

Assuming that all of the on-site energies are equal, the Hamiltonian $H_e$ of the central cluster has six eigenenergies at $E = \epsilon \pm 2t$ and $E = \epsilon \pm t$ which the latter is 2-fold degenerate. For this example we set the on-site energy, $\epsilon_i=0$, and hopping parameter, $t=-3 eV$. The calculation of the transmission, $T$, in the weak coupling regime is shown in Fig. (\ref{fig2}).

\begin{figure}[!h]
	\centering
	\includegraphics[width=0.5\textwidth]{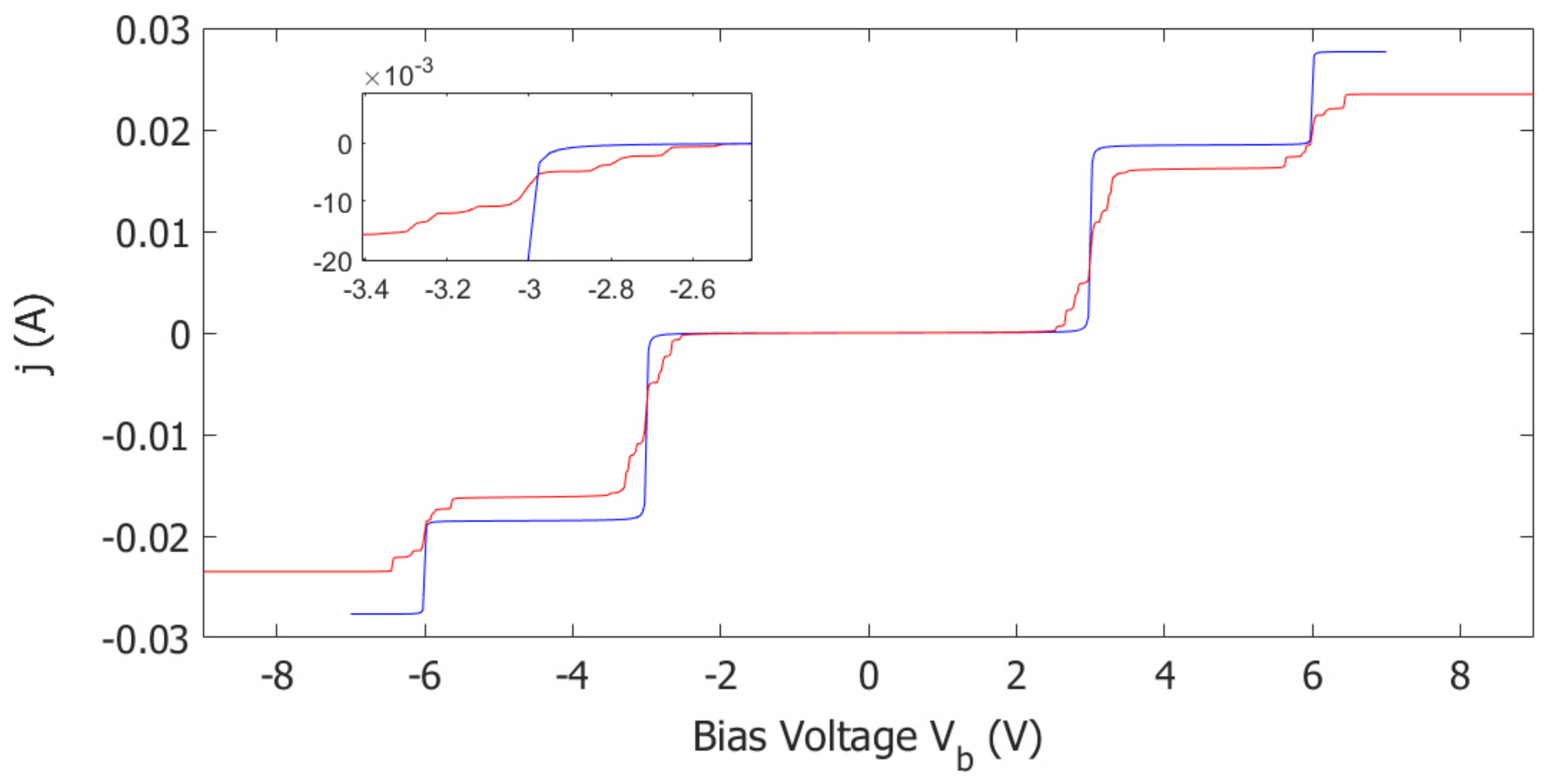}
	\caption{Illustration of the current versus bias voltage $V_b$ in the absence of electron-phonon interaction (blue) and in the presence of the interaction (red). The effect of electron-phonon interaction on the current is visible. Then inset shows the current at $V_b=3V$.}
	\label{fig3}
\end{figure}

The four expected peaks centered at the eigenenergies of the system show the impact of electron-phonon interaction. Increasing the coupling constant to leads, $\tau$, the visibility of these small peaks declines and they merge together. The current evaluated in this regime is depicted in Fig. (\ref{fig3}). In the absence of electron-phonon interaction, the current profile of the system shows 4 main steps. The impact of interaction is to divide each of them into smaller steps. The number of these steps is determined according to the number of ways a phonon can be absorbed or emitted. The current increases stepwise with bias voltage at voltage differences proportional to phonon energies $\hbar \omega_{q_i}$ and electron-phonon coupling constant. 

Similar to the argument about the transmission, here also increasing the tunneling coupling constant $\tau$ will lead to washing out the steps in current.

Same results have been found in \cite{PhysRevB.88.054301} but for the case of breathing mode in a Benzene molecule.

\begin{figure}[!h]
	\centering
	\includegraphics[width=0.48\textwidth]{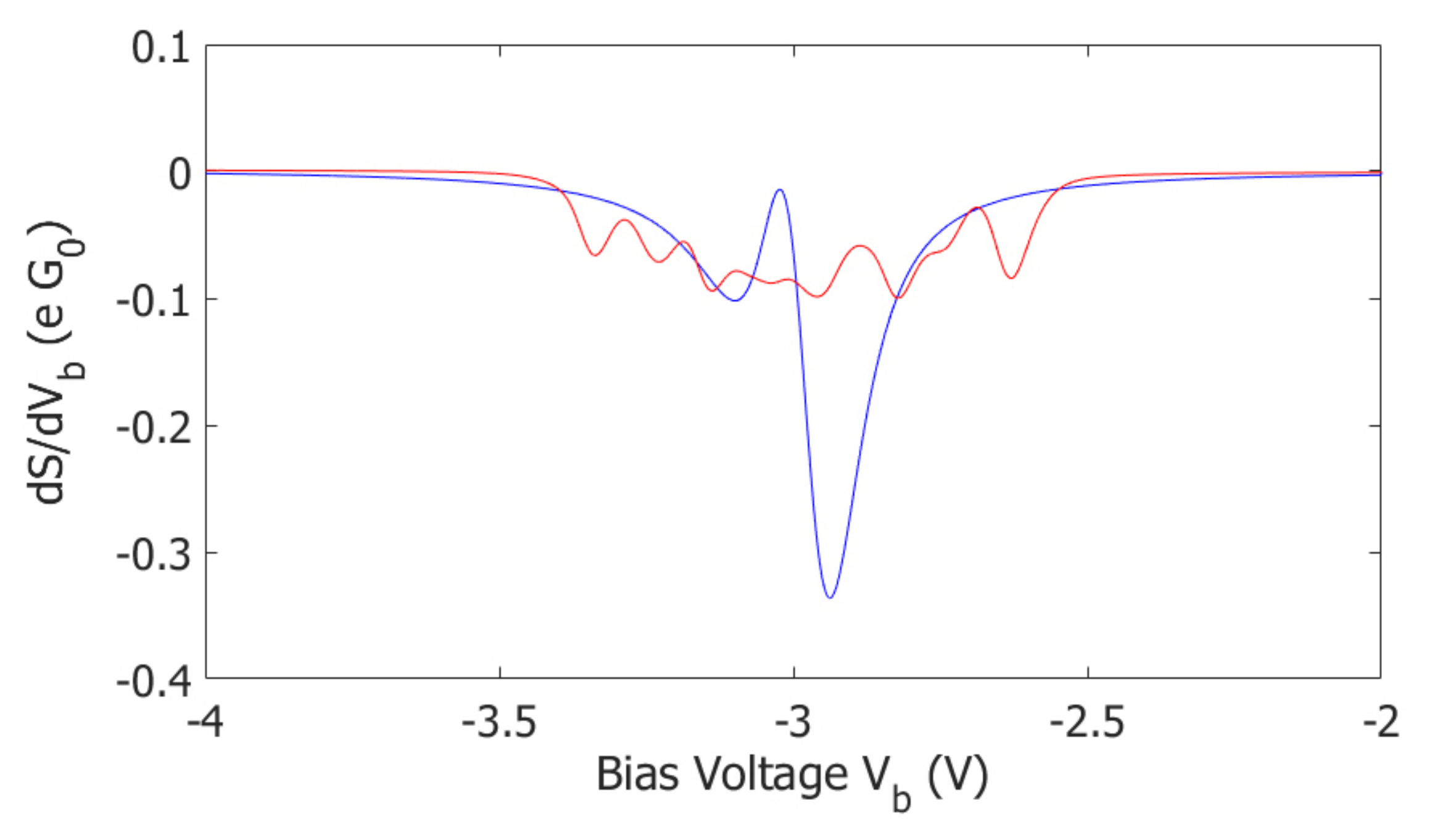}
	\caption{Comparison of the differential shot noise around bias voltage $V_B = -3 eV$ in units of $e^2/h$ with (red) and without (blue) the interactions. The phonon-mediated peaks are visible.}
	\label{fig4}
\end{figure}

Finally, we study the effect of electron-phonon interaction on the shot noise. Correspondingly, one can see peaks in differential shot noise, (\ref{fig4}). They are located around where the conductance peaked. As stated in \cite{PhysRevB.67.165326}, the reason for this behavior is that no noise is generated when the transmission probability is 0 or 1, while it is non-zero in between.
\section{Summary}
We have explored the effect of electron-phonon interaction on the transport properties of an $N$-atomic periodic chain. The main idea was to avoid local electron-phonon interaction or Einstein-phonon assumption. We have used the Green's function formalism together with Fr\"{o}hlich Hamiltonian and studied the model using the exact diagonalization technique. Finally we applied our method to special case of $N=6$ which can be regarded as benzene-like molecule.

\bibliography{bibliog}

\begin{thebibliography}{40}%
\makeatletter
\providecommand \@ifxundefined [1]{%
 \@ifx{#1\undefined}
}%
\providecommand \@ifnum [1]{%
 \ifnum #1\expandafter \@firstoftwo
 \else \expandafter \@secondoftwo
 \fi
}%
\providecommand \@ifx [1]{%
 \ifx #1\expandafter \@firstoftwo
 \else \expandafter \@secondoftwo
 \fi
}%
\providecommand \natexlab [1]{#1}%
\providecommand \enquote  [1]{``#1''}%
\providecommand \bibnamefont  [1]{#1}%
\providecommand \bibfnamefont [1]{#1}%
\providecommand \citenamefont [1]{#1}%
\providecommand \href@noop [0]{\@secondoftwo}%
\providecommand \href [0]{\begingroup \@sanitize@url \@href}%
\providecommand \@href[1]{\@@startlink{#1}\@@href}%
\providecommand \@@href[1]{\endgroup#1\@@endlink}%
\providecommand \@sanitize@url [0]{\catcode `\\12\catcode `\$12\catcode
  `\&12\catcode `\#12\catcode `\^12\catcode `\_12\catcode `\%12\relax}%
\providecommand \@@startlink[1]{}%
\providecommand \@@endlink[0]{}%
\providecommand \url  [0]{\begingroup\@sanitize@url \@url }%
\providecommand \@url [1]{\endgroup\@href {#1}{\urlprefix }}%
\providecommand \urlprefix  [0]{URL }%
\providecommand \Eprint [0]{\href }%
\providecommand \doibase [0]{http://dx.doi.org/}%
\providecommand \selectlanguage [0]{\@gobble}%
\providecommand \bibinfo  [0]{\@secondoftwo}%
\providecommand \bibfield  [0]{\@secondoftwo}%
\providecommand \translation [1]{[#1]}%
\providecommand \BibitemOpen [0]{}%
\providecommand \bibitemStop [0]{}%
\providecommand \bibitemNoStop [0]{.\EOS\space}%
\providecommand \EOS [0]{\spacefactor3000\relax}%
\providecommand \BibitemShut  [1]{\csname bibitem#1\endcsname}%
\let\auto@bib@innerbib\@empty
\bibitem [{\citenamefont {Bardeen}\ \emph {et~al.}(1957)\citenamefont
  {Bardeen}, \citenamefont {Cooper},\ and\ \citenamefont
  {Schrieffer}}]{PhysRev.108.1175}%
  \BibitemOpen
  \bibfield  {author} {\bibinfo {author} {\bibfnamefont {J.}~\bibnamefont
  {Bardeen}}, \bibinfo {author} {\bibfnamefont {L.~N.}\ \bibnamefont {Cooper}},
  \ and\ \bibinfo {author} {\bibfnamefont {J.~R.}\ \bibnamefont {Schrieffer}},\
  }\href {\doibase 10.1103/PhysRev.108.1175} {\bibfield  {journal} {\bibinfo
  {journal} {Phys. Rev.}\ }\textbf {\bibinfo {volume} {108}},\ \bibinfo {pages}
  {1175} (\bibinfo {year} {1957})}\BibitemShut {NoStop}%
\bibitem [{\citenamefont {Delaire}\ \emph {et~al.}(2011)\citenamefont
  {Delaire}, \citenamefont {Ma}, \citenamefont {Marty}, \citenamefont {May},
  \citenamefont {McGuire}, \citenamefont {Du}, \citenamefont {Singh},
  \citenamefont {Podlesnyak}, \citenamefont {Ehlers}, \citenamefont {Lumsden},\
  and\ \citenamefont {Sales}}]{Delaire2011}%
  \BibitemOpen
  \bibfield  {author} {\bibinfo {author} {\bibfnamefont {O.}~\bibnamefont
  {Delaire}}, \bibinfo {author} {\bibfnamefont {J.}~\bibnamefont {Ma}},
  \bibinfo {author} {\bibfnamefont {K.}~\bibnamefont {Marty}}, \bibinfo
  {author} {\bibfnamefont {A.~F.}\ \bibnamefont {May}}, \bibinfo {author}
  {\bibfnamefont {M.~A.}\ \bibnamefont {McGuire}}, \bibinfo {author}
  {\bibfnamefont {M.-H.}\ \bibnamefont {Du}}, \bibinfo {author} {\bibfnamefont
  {D.~J.}\ \bibnamefont {Singh}}, \bibinfo {author} {\bibfnamefont
  {A.}~\bibnamefont {Podlesnyak}}, \bibinfo {author} {\bibfnamefont
  {G.}~\bibnamefont {Ehlers}}, \bibinfo {author} {\bibfnamefont {M.~D.}\
  \bibnamefont {Lumsden}}, \ and\ \bibinfo {author} {\bibfnamefont {B.~C.}\
  \bibnamefont {Sales}},\ }\href {\doibase 10.1038/nmat3035} {\bibfield
  {journal} {\bibinfo  {journal} {Nat Mater}\ }\textbf {\bibinfo {volume}
  {10}},\ \bibinfo {pages} {614} (\bibinfo {year} {2011})}\BibitemShut
  {NoStop}%
\bibitem [{\citenamefont {Millis}(1998)}]{1473}%
  \BibitemOpen
  \bibfield  {author} {\bibinfo {author} {\bibfnamefont {A.~J.}\ \bibnamefont
  {Millis}},\ }\href {\doibase 10.1098/rsta.1998.0230} {\bibfield  {journal}
  {\bibinfo  {journal} {Philos. Trans. R. Soc. London, Ser. A}\ }\textbf
  {\bibinfo {volume} {356}},\ \bibinfo {pages} {1473} (\bibinfo {year}
  {1998})}\BibitemShut {NoStop}%
\bibitem [{\citenamefont {Flensberg}(2003)}]{PhysRevB.68.205323}%
  \BibitemOpen
  \bibfield  {author} {\bibinfo {author} {\bibfnamefont {K.}~\bibnamefont
  {Flensberg}},\ }\href {\doibase 10.1103/PhysRevB.68.205323} {\bibfield
  {journal} {\bibinfo  {journal} {Phys. Rev. B}\ }\textbf {\bibinfo {volume}
  {68}},\ \bibinfo {pages} {205323} (\bibinfo {year} {2003})}\BibitemShut
  {NoStop}%
\bibitem [{\citenamefont {Zazunov}\ \emph {et~al.}(2006)\citenamefont
  {Zazunov}, \citenamefont {Feinberg},\ and\ \citenamefont
  {Martin}}]{PhysRevB.73.115405}%
  \BibitemOpen
  \bibfield  {author} {\bibinfo {author} {\bibfnamefont {A.}~\bibnamefont
  {Zazunov}}, \bibinfo {author} {\bibfnamefont {D.}~\bibnamefont {Feinberg}}, \
  and\ \bibinfo {author} {\bibfnamefont {T.}~\bibnamefont {Martin}},\ }\href
  {\doibase 10.1103/PhysRevB.73.115405} {\bibfield  {journal} {\bibinfo
  {journal} {Phys. Rev. B}\ }\textbf {\bibinfo {volume} {73}},\ \bibinfo
  {pages} {115405} (\bibinfo {year} {2006})}\BibitemShut {NoStop}%
\bibitem [{\citenamefont {Koch}\ and\ \citenamefont {von
  Oppen}(2005)}]{PhysRevLett.94.206804}%
  \BibitemOpen
  \bibfield  {author} {\bibinfo {author} {\bibfnamefont {J.}~\bibnamefont
  {Koch}}\ and\ \bibinfo {author} {\bibfnamefont {F.}~\bibnamefont {von
  Oppen}},\ }\href {\doibase 10.1103/PhysRevLett.94.206804} {\bibfield
  {journal} {\bibinfo  {journal} {Phys. Rev. Lett.}\ }\textbf {\bibinfo
  {volume} {94}},\ \bibinfo {pages} {206804} (\bibinfo {year}
  {2005})}\BibitemShut {NoStop}%
\bibitem [{\citenamefont {Galperin}\ \emph {et~al.}(2006)\citenamefont
  {Galperin}, \citenamefont {Nitzan},\ and\ \citenamefont
  {Ratner}}]{PhysRevB.73.045314}%
  \BibitemOpen
  \bibfield  {author} {\bibinfo {author} {\bibfnamefont {M.}~\bibnamefont
  {Galperin}}, \bibinfo {author} {\bibfnamefont {A.}~\bibnamefont {Nitzan}}, \
  and\ \bibinfo {author} {\bibfnamefont {M.~A.}\ \bibnamefont {Ratner}},\
  }\href {\doibase 10.1103/PhysRevB.73.045314} {\bibfield  {journal} {\bibinfo
  {journal} {Phys. Rev. B}\ }\textbf {\bibinfo {volume} {73}},\ \bibinfo
  {pages} {045314} (\bibinfo {year} {2006})}\BibitemShut {NoStop}%
\bibitem [{\citenamefont {Benesch}\ \emph {et~al.}(2006)\citenamefont
  {Benesch}, \citenamefont {\v{C}\'{i}\v{z}ek}, \citenamefont {Thoss},\ and\
  \citenamefont {Domcke}}]{Benesch2006355}%
  \BibitemOpen
  \bibfield  {author} {\bibinfo {author} {\bibfnamefont {C.}~\bibnamefont
  {Benesch}}, \bibinfo {author} {\bibfnamefont {M.}~\bibnamefont
  {\v{C}\'{i}\v{z}ek}}, \bibinfo {author} {\bibfnamefont {M.}~\bibnamefont
  {Thoss}}, \ and\ \bibinfo {author} {\bibfnamefont {W.}~\bibnamefont
  {Domcke}},\ }\href {\doibase http://dx.doi.org/10.1016/j.cplett.2006.09.003}
  {\bibfield  {journal} {\bibinfo  {journal} {Chemical Physics Letters}\
  }\textbf {\bibinfo {volume} {430}},\ \bibinfo {pages} {355 } (\bibinfo {year}
  {2006})}\BibitemShut {NoStop}%
\bibitem [{\citenamefont {Ryndyk}\ and\ \citenamefont
  {Cuniberti}(2007)}]{PhysRevB.76.155430}%
  \BibitemOpen
  \bibfield  {author} {\bibinfo {author} {\bibfnamefont {D.~A.}\ \bibnamefont
  {Ryndyk}}\ and\ \bibinfo {author} {\bibfnamefont {G.}~\bibnamefont
  {Cuniberti}},\ }\href {\doibase 10.1103/PhysRevB.76.155430} {\bibfield
  {journal} {\bibinfo  {journal} {Phys. Rev. B}\ }\textbf {\bibinfo {volume}
  {76}},\ \bibinfo {pages} {155430} (\bibinfo {year} {2007})}\BibitemShut
  {NoStop}%
\bibitem [{\citenamefont {Galperin}\ \emph {et~al.}(2007)\citenamefont
  {Galperin}, \citenamefont {Ratner},\ and\ \citenamefont
  {Nitzan}}]{0953-8984-19-10-103201}%
  \BibitemOpen
  \bibfield  {author} {\bibinfo {author} {\bibfnamefont {M.}~\bibnamefont
  {Galperin}}, \bibinfo {author} {\bibfnamefont {M.~A.}\ \bibnamefont
  {Ratner}}, \ and\ \bibinfo {author} {\bibfnamefont {A.}~\bibnamefont
  {Nitzan}},\ }\href {http://stacks.iop.org/0953-8984/19/i=10/a=103201}
  {\bibfield  {journal} {\bibinfo  {journal} {Journal of Physics: Condensed
  Matter}\ }\textbf {\bibinfo {volume} {19}},\ \bibinfo {pages} {103201}
  (\bibinfo {year} {2007})}\BibitemShut {NoStop}%
\bibitem [{\citenamefont {Zhitenev}\ \emph {et~al.}(2002)\citenamefont
  {Zhitenev}, \citenamefont {Meng},\ and\ \citenamefont
  {Bao}}]{PhysRevLett.88.226801}%
  \BibitemOpen
  \bibfield  {author} {\bibinfo {author} {\bibfnamefont {N.~B.}\ \bibnamefont
  {Zhitenev}}, \bibinfo {author} {\bibfnamefont {H.}~\bibnamefont {Meng}}, \
  and\ \bibinfo {author} {\bibfnamefont {Z.}~\bibnamefont {Bao}},\ }\href
  {\doibase 10.1103/PhysRevLett.88.226801} {\bibfield  {journal} {\bibinfo
  {journal} {Phys. Rev. Lett.}\ }\textbf {\bibinfo {volume} {88}},\ \bibinfo
  {pages} {226801} (\bibinfo {year} {2002})}\BibitemShut {NoStop}%
\bibitem [{\citenamefont {LeRoy}\ \emph {et~al.}(2005)\citenamefont {LeRoy},
  \citenamefont {Kong}, \citenamefont {Pahilwani}, \citenamefont {Dekker},\
  and\ \citenamefont {Lemay}}]{PhysRevB.72.075413}%
  \BibitemOpen
  \bibfield  {author} {\bibinfo {author} {\bibfnamefont {B.~J.}\ \bibnamefont
  {LeRoy}}, \bibinfo {author} {\bibfnamefont {J.}~\bibnamefont {Kong}},
  \bibinfo {author} {\bibfnamefont {V.~K.}\ \bibnamefont {Pahilwani}}, \bibinfo
  {author} {\bibfnamefont {C.}~\bibnamefont {Dekker}}, \ and\ \bibinfo {author}
  {\bibfnamefont {S.~G.}\ \bibnamefont {Lemay}},\ }\href {\doibase
  10.1103/PhysRevB.72.075413} {\bibfield  {journal} {\bibinfo  {journal} {Phys.
  Rev. B}\ }\textbf {\bibinfo {volume} {72}},\ \bibinfo {pages} {075413}
  (\bibinfo {year} {2005})}\BibitemShut {NoStop}%
\bibitem [{\citenamefont {Qiu}\ \emph {et~al.}(2004)\citenamefont {Qiu},
  \citenamefont {Nazin},\ and\ \citenamefont {Ho}}]{PhysRevLett.92.206102}%
  \BibitemOpen
  \bibfield  {author} {\bibinfo {author} {\bibfnamefont {X.~H.}\ \bibnamefont
  {Qiu}}, \bibinfo {author} {\bibfnamefont {G.~V.}\ \bibnamefont {Nazin}}, \
  and\ \bibinfo {author} {\bibfnamefont {W.}~\bibnamefont {Ho}},\ }\href
  {\doibase 10.1103/PhysRevLett.92.206102} {\bibfield  {journal} {\bibinfo
  {journal} {Phys. Rev. Lett.}\ }\textbf {\bibinfo {volume} {92}},\ \bibinfo
  {pages} {206102} (\bibinfo {year} {2004})}\BibitemShut {NoStop}%
\bibitem [{\citenamefont {Park}\ \emph {et~al.}(2000)\citenamefont {Park},
  \citenamefont {Park}, \citenamefont {Lim}, \citenamefont {Anderson},
  \citenamefont {Alivisatos},\ and\ \citenamefont {McEuen}}]{Park2000}%
  \BibitemOpen
  \bibfield  {author} {\bibinfo {author} {\bibfnamefont {H.}~\bibnamefont
  {Park}}, \bibinfo {author} {\bibfnamefont {J.}~\bibnamefont {Park}}, \bibinfo
  {author} {\bibfnamefont {A.~K.~L.}\ \bibnamefont {Lim}}, \bibinfo {author}
  {\bibfnamefont {E.~H.}\ \bibnamefont {Anderson}}, \bibinfo {author}
  {\bibfnamefont {A.~P.}\ \bibnamefont {Alivisatos}}, \ and\ \bibinfo {author}
  {\bibfnamefont {P.~L.}\ \bibnamefont {McEuen}},\ }\href {\doibase
  10.1038/35024031} {\bibfield  {journal} {\bibinfo  {journal} {Nature}\
  }\textbf {\bibinfo {volume} {407}},\ \bibinfo {pages} {57} (\bibinfo {year}
  {2000})}\BibitemShut {NoStop}%
\bibitem [{\citenamefont {Marsiglio}(1993)}]{MARSIGLIO1993280}%
  \BibitemOpen
  \bibfield  {author} {\bibinfo {author} {\bibfnamefont {F.}~\bibnamefont
  {Marsiglio}},\ }\href {\doibase
  http://dx.doi.org/10.1016/0375-9601(93)90711-8} {\bibfield  {journal}
  {\bibinfo  {journal} {Physics Letters A}\ }\textbf {\bibinfo {volume}
  {180}},\ \bibinfo {pages} {280 } (\bibinfo {year} {1993})}\BibitemShut
  {NoStop}%
\bibitem [{\citenamefont {Nath}\ \emph {et~al.}(2015)\citenamefont {Nath},
  \citenamefont {Mondal},\ and\ \citenamefont {Ghosh}}]{Nath2015}%
  \BibitemOpen
  \bibfield  {author} {\bibinfo {author} {\bibfnamefont {S.}~\bibnamefont
  {Nath}}, \bibinfo {author} {\bibfnamefont {N.~S.}\ \bibnamefont {Mondal}}, \
  and\ \bibinfo {author} {\bibfnamefont {N.~K.}\ \bibnamefont {Ghosh}},\ }\href
  {\doibase 10.1007/s10948-015-2957-1} {\bibfield  {journal} {\bibinfo
  {journal} {Journal of Superconductivity and Novel Magnetism}\ }\textbf
  {\bibinfo {volume} {28}},\ \bibinfo {pages} {1687} (\bibinfo {year}
  {2015})}\BibitemShut {NoStop}%
\bibitem [{\citenamefont {de~Mello}\ and\ \citenamefont
  {Ranninger}(1997)}]{PhysRevB.55.14872}%
  \BibitemOpen
  \bibfield  {author} {\bibinfo {author} {\bibfnamefont {E.~V.~L.}\
  \bibnamefont {de~Mello}}\ and\ \bibinfo {author} {\bibfnamefont
  {J.}~\bibnamefont {Ranninger}},\ }\href {\doibase 10.1103/PhysRevB.55.14872}
  {\bibfield  {journal} {\bibinfo  {journal} {Phys. Rev. B}\ }\textbf {\bibinfo
  {volume} {55}},\ \bibinfo {pages} {14872} (\bibinfo {year}
  {1997})}\BibitemShut {NoStop}%
\bibitem [{\citenamefont {Karakonstantakis}\ \emph {et~al.}(2013)\citenamefont
  {Karakonstantakis}, \citenamefont {Liu}, \citenamefont {Thomale},\ and\
  \citenamefont {Kivelson}}]{PhysRevB.88.224512}%
  \BibitemOpen
  \bibfield  {author} {\bibinfo {author} {\bibfnamefont {G.}~\bibnamefont
  {Karakonstantakis}}, \bibinfo {author} {\bibfnamefont {L.}~\bibnamefont
  {Liu}}, \bibinfo {author} {\bibfnamefont {R.}~\bibnamefont {Thomale}}, \ and\
  \bibinfo {author} {\bibfnamefont {S.~A.}\ \bibnamefont {Kivelson}},\ }\href
  {\doibase 10.1103/PhysRevB.88.224512} {\bibfield  {journal} {\bibinfo
  {journal} {Phys. Rev. B}\ }\textbf {\bibinfo {volume} {88}},\ \bibinfo
  {pages} {224512} (\bibinfo {year} {2013})}\BibitemShut {NoStop}%
\bibitem [{\citenamefont {Tezuka}\ \emph {et~al.}(2005)\citenamefont {Tezuka},
  \citenamefont {Arita},\ and\ \citenamefont {Aoki}}]{Tezuka2005708}%
  \BibitemOpen
  \bibfield  {author} {\bibinfo {author} {\bibfnamefont {M.}~\bibnamefont
  {Tezuka}}, \bibinfo {author} {\bibfnamefont {R.}~\bibnamefont {Arita}}, \
  and\ \bibinfo {author} {\bibfnamefont {H.}~\bibnamefont {Aoki}},\ }\href
  {\doibase http://dx.doi.org/10.1016/j.physb.2005.01.201} {\bibfield
  {journal} {\bibinfo  {journal} {Physica B: Condensed Matter}\ }\textbf
  {\bibinfo {volume} {359–361}},\ \bibinfo {pages} {708 } (\bibinfo {year}
  {2005})},\ \bibinfo {note} {proceedings of the International Conference on
  Strongly Correlated Electron Systems}\BibitemShut {NoStop}%
\bibitem [{\citenamefont {Tozer}\ and\ \citenamefont
  {Barford}(2014)}]{PhysRevB.89.155434}%
  \BibitemOpen
  \bibfield  {author} {\bibinfo {author} {\bibfnamefont {O.~R.}\ \bibnamefont
  {Tozer}}\ and\ \bibinfo {author} {\bibfnamefont {W.}~\bibnamefont
  {Barford}},\ }\href {\doibase 10.1103/PhysRevB.89.155434} {\bibfield
  {journal} {\bibinfo  {journal} {Phys. Rev. B}\ }\textbf {\bibinfo {volume}
  {89}},\ \bibinfo {pages} {155434} (\bibinfo {year} {2014})}\BibitemShut
  {NoStop}%
\bibitem [{\citenamefont {Hirsch}(1983)}]{PhysRevLett.51.296}%
  \BibitemOpen
  \bibfield  {author} {\bibinfo {author} {\bibfnamefont {J.~E.}\ \bibnamefont
  {Hirsch}},\ }\href {\doibase 10.1103/PhysRevLett.51.296} {\bibfield
  {journal} {\bibinfo  {journal} {Phys. Rev. Lett.}\ }\textbf {\bibinfo
  {volume} {51}},\ \bibinfo {pages} {296} (\bibinfo {year} {1983})}\BibitemShut
  {NoStop}%
\bibitem [{\citenamefont {Mason}\ and\ \citenamefont
  {Hess}(1989)}]{PhysRevB.39.5051}%
  \BibitemOpen
  \bibfield  {author} {\bibinfo {author} {\bibfnamefont {B.~A.}\ \bibnamefont
  {Mason}}\ and\ \bibinfo {author} {\bibfnamefont {K.}~\bibnamefont {Hess}},\
  }\href {\doibase 10.1103/PhysRevB.39.5051} {\bibfield  {journal} {\bibinfo
  {journal} {Phys. Rev. B}\ }\textbf {\bibinfo {volume} {39}},\ \bibinfo
  {pages} {5051} (\bibinfo {year} {1989})}\BibitemShut {NoStop}%
\bibitem [{\citenamefont {Ohgoe}\ and\ \citenamefont
  {Imada}(2014)}]{PhysRevB.89.195139}%
  \BibitemOpen
  \bibfield  {author} {\bibinfo {author} {\bibfnamefont {T.}~\bibnamefont
  {Ohgoe}}\ and\ \bibinfo {author} {\bibfnamefont {M.}~\bibnamefont {Imada}},\
  }\href {\doibase 10.1103/PhysRevB.89.195139} {\bibfield  {journal} {\bibinfo
  {journal} {Phys. Rev. B}\ }\textbf {\bibinfo {volume} {89}},\ \bibinfo
  {pages} {195139} (\bibinfo {year} {2014})}\BibitemShut {NoStop}%
\bibitem [{\citenamefont {Kuchinskii}\ \emph {et~al.}(2012)\citenamefont
  {Kuchinskii}, \citenamefont {Nekrasov},\ and\ \citenamefont
  {Sadovskii}}]{1063-7869-55-4-R01}%
  \BibitemOpen
  \bibfield  {author} {\bibinfo {author} {\bibfnamefont {E.~Z.}\ \bibnamefont
  {Kuchinskii}}, \bibinfo {author} {\bibfnamefont {I.~A.}\ \bibnamefont
  {Nekrasov}}, \ and\ \bibinfo {author} {\bibfnamefont {M.~V.}\ \bibnamefont
  {Sadovskii}},\ }\href {http://stacks.iop.org/1063-7869/55/i=4/a=R01}
  {\bibfield  {journal} {\bibinfo  {journal} {Physics-Uspekhi}\ }\textbf
  {\bibinfo {volume} {55}},\ \bibinfo {pages} {325} (\bibinfo {year}
  {2012})}\BibitemShut {NoStop}%
\bibitem [{\citenamefont {Matsueda}\ \emph {et~al.}(2008)\citenamefont
  {Matsueda}, \citenamefont {Ando}, \citenamefont {Tohyama},\ and\
  \citenamefont {Maekawa}}]{Matsueda20083070}%
  \BibitemOpen
  \bibfield  {author} {\bibinfo {author} {\bibfnamefont {H.}~\bibnamefont
  {Matsueda}}, \bibinfo {author} {\bibfnamefont {A.}~\bibnamefont {Ando}},
  \bibinfo {author} {\bibfnamefont {T.}~\bibnamefont {Tohyama}}, \ and\
  \bibinfo {author} {\bibfnamefont {S.}~\bibnamefont {Maekawa}},\ }\href
  {\doibase http://dx.doi.org/10.1016/j.jpcs.2008.06.040} {\bibfield  {journal}
  {\bibinfo  {journal} {Journal of Physics and Chemistry of Solids}\ }\textbf
  {\bibinfo {volume} {69}},\ \bibinfo {pages} {3070 } (\bibinfo {year}
  {2008})},\ \bibinfo {note} {\{SNS2007Spectroscopies\} in Novel
  Superconductors '07}\BibitemShut {NoStop}%
\bibitem [{\citenamefont {Holstein}(1959)}]{HOLSTEIN1959325}%
  \BibitemOpen
  \bibfield  {author} {\bibinfo {author} {\bibfnamefont {T.}~\bibnamefont
  {Holstein}},\ }\href {\doibase
  http://dx.doi.org/10.1016/0003-4916(59)90002-8} {\bibfield  {journal}
  {\bibinfo  {journal} {Annals of Physics}\ }\textbf {\bibinfo {volume} {8}},\
  \bibinfo {pages} {325 } (\bibinfo {year} {1959})}\BibitemShut {NoStop}%
\bibitem [{\citenamefont {Lee}\ \emph {et~al.}(2009)\citenamefont {Lee},
  \citenamefont {Jean},\ and\ \citenamefont {Sanvito}}]{PhysRevB.79.085120}%
  \BibitemOpen
  \bibfield  {author} {\bibinfo {author} {\bibfnamefont {W.}~\bibnamefont
  {Lee}}, \bibinfo {author} {\bibfnamefont {N.}~\bibnamefont {Jean}}, \ and\
  \bibinfo {author} {\bibfnamefont {S.}~\bibnamefont {Sanvito}},\ }\href
  {\doibase 10.1103/PhysRevB.79.085120} {\bibfield  {journal} {\bibinfo
  {journal} {Phys. Rev. B}\ }\textbf {\bibinfo {volume} {79}},\ \bibinfo
  {pages} {085120} (\bibinfo {year} {2009})}\BibitemShut {NoStop}%
\bibitem [{\citenamefont {Devreese}\ and\ \citenamefont
  {Alexandrov}(2009)}]{frohlichpolaron}%
  \BibitemOpen
  \bibfield  {author} {\bibinfo {author} {\bibfnamefont {J.~T.}\ \bibnamefont
  {Devreese}}\ and\ \bibinfo {author} {\bibfnamefont {A.~S.}\ \bibnamefont
  {Alexandrov}},\ }\href {http://stacks.iop.org/0034-4885/72/i=6/a=066501}
  {\bibfield  {journal} {\bibinfo  {journal} {Reports on Progress in Physics}\
  }\textbf {\bibinfo {volume} {72}},\ \bibinfo {pages} {066501} (\bibinfo
  {year} {2009})}\BibitemShut {NoStop}%
\bibitem [{\citenamefont {Georges}\ \emph {et~al.}(1996)\citenamefont
  {Georges}, \citenamefont {Kotliar}, \citenamefont {Krauth},\ and\
  \citenamefont {Rozenberg}}]{RevModPhys.68.13}%
  \BibitemOpen
  \bibfield  {author} {\bibinfo {author} {\bibfnamefont {A.}~\bibnamefont
  {Georges}}, \bibinfo {author} {\bibfnamefont {G.}~\bibnamefont {Kotliar}},
  \bibinfo {author} {\bibfnamefont {W.}~\bibnamefont {Krauth}}, \ and\ \bibinfo
  {author} {\bibfnamefont {M.~J.}\ \bibnamefont {Rozenberg}},\ }\href {\doibase
  10.1103/RevModPhys.68.13} {\bibfield  {journal} {\bibinfo  {journal} {Rev.
  Mod. Phys.}\ }\textbf {\bibinfo {volume} {68}},\ \bibinfo {pages} {13}
  (\bibinfo {year} {1996})}\BibitemShut {NoStop}%
\bibitem [{\citenamefont {Song}\ \emph {et~al.}(2009)\citenamefont {Song},
  \citenamefont {Kim}, \citenamefont {Jang}, \citenamefont {Jeong},
  \citenamefont {Reed},\ and\ \citenamefont {Lee}}]{Song2009}%
  \BibitemOpen
  \bibfield  {author} {\bibinfo {author} {\bibfnamefont {H.}~\bibnamefont
  {Song}}, \bibinfo {author} {\bibfnamefont {Y.}~\bibnamefont {Kim}}, \bibinfo
  {author} {\bibfnamefont {Y.~H.}\ \bibnamefont {Jang}}, \bibinfo {author}
  {\bibfnamefont {H.}~\bibnamefont {Jeong}}, \bibinfo {author} {\bibfnamefont
  {M.~A.}\ \bibnamefont {Reed}}, \ and\ \bibinfo {author} {\bibfnamefont
  {T.}~\bibnamefont {Lee}},\ }\href {\doibase 10.1038/nature08639} {\bibfield
  {journal} {\bibinfo  {journal} {Nature}\ }\textbf {\bibinfo {volume} {462}},\
  \bibinfo {pages} {1039} (\bibinfo {year} {2009})}\BibitemShut {NoStop}%
\bibitem [{\citenamefont {Mondal}\ and\ \citenamefont
  {Ghosh}(2011)}]{Mondal20113723}%
  \BibitemOpen
  \bibfield  {author} {\bibinfo {author} {\bibfnamefont {N.}~\bibnamefont
  {Mondal}}\ and\ \bibinfo {author} {\bibfnamefont {N.}~\bibnamefont {Ghosh}},\
  }\href {\doibase http://dx.doi.org/10.1016/j.physb.2011.07.002} {\bibfield
  {journal} {\bibinfo  {journal} {Physica B: Condensed Matter}\ }\textbf
  {\bibinfo {volume} {406}},\ \bibinfo {pages} {3723 } (\bibinfo {year}
  {2011})}\BibitemShut {NoStop}%
\bibitem [{\citenamefont {Beugeling}\ \emph {et~al.}(2014)\citenamefont
  {Beugeling}, \citenamefont {Moessner},\ and\ \citenamefont
  {Haque}}]{PhysRevE.89.042112}%
  \BibitemOpen
  \bibfield  {author} {\bibinfo {author} {\bibfnamefont {W.}~\bibnamefont
  {Beugeling}}, \bibinfo {author} {\bibfnamefont {R.}~\bibnamefont {Moessner}},
  \ and\ \bibinfo {author} {\bibfnamefont {M.}~\bibnamefont {Haque}},\ }\href
  {\doibase 10.1103/PhysRevE.89.042112} {\bibfield  {journal} {\bibinfo
  {journal} {Phys. Rev. E}\ }\textbf {\bibinfo {volume} {89}},\ \bibinfo
  {pages} {042112} (\bibinfo {year} {2014})}\BibitemShut {NoStop}%
\bibitem [{\citenamefont {Zhang}\ \emph {et~al.}(1998)\citenamefont {Zhang},
  \citenamefont {Jeckelmann},\ and\ \citenamefont
  {White}}]{PhysRevLett.80.2661}%
  \BibitemOpen
  \bibfield  {author} {\bibinfo {author} {\bibfnamefont {C.}~\bibnamefont
  {Zhang}}, \bibinfo {author} {\bibfnamefont {E.}~\bibnamefont {Jeckelmann}}, \
  and\ \bibinfo {author} {\bibfnamefont {S.~R.}\ \bibnamefont {White}},\ }\href
  {\doibase 10.1103/PhysRevLett.80.2661} {\bibfield  {journal} {\bibinfo
  {journal} {Phys. Rev. Lett.}\ }\textbf {\bibinfo {volume} {80}},\ \bibinfo
  {pages} {2661} (\bibinfo {year} {1998})}\BibitemShut {NoStop}%
\bibitem [{\citenamefont {Zhang}\ \emph {et~al.}(1999)\citenamefont {Zhang},
  \citenamefont {Jeckelmann},\ and\ \citenamefont {White}}]{PhysRevB.60.14092}%
  \BibitemOpen
  \bibfield  {author} {\bibinfo {author} {\bibfnamefont {C.}~\bibnamefont
  {Zhang}}, \bibinfo {author} {\bibfnamefont {E.}~\bibnamefont {Jeckelmann}}, \
  and\ \bibinfo {author} {\bibfnamefont {S.~R.}\ \bibnamefont {White}},\ }\href
  {\doibase 10.1103/PhysRevB.60.14092} {\bibfield  {journal} {\bibinfo
  {journal} {Phys. Rev. B}\ }\textbf {\bibinfo {volume} {60}},\ \bibinfo
  {pages} {14092} (\bibinfo {year} {1999})}\BibitemShut {NoStop}%
\bibitem [{\citenamefont {Cuevas}\ and\ \citenamefont
  {Scheer}()}]{MolecElectrnoic}%
  \BibitemOpen
  \bibfield  {author} {\bibinfo {author} {\bibfnamefont {J.~C.}\ \bibnamefont
  {Cuevas}}\ and\ \bibinfo {author} {\bibfnamefont {E.}~\bibnamefont
  {Scheer}},\ }\href@noop {} {\emph {\bibinfo {title} {Molecular Electronics -
  An Introduction to Theory and Experiment}}}\BibitemShut {NoStop}%
\bibitem [{\citenamefont {S\'en\'echal}\ \emph {et~al.}(2000)\citenamefont
  {S\'en\'echal}, \citenamefont {Perez},\ and\ \citenamefont
  {Pioro-Ladri\`ere}}]{PhysRevLett.84.522}%
  \BibitemOpen
  \bibfield  {author} {\bibinfo {author} {\bibfnamefont {D.}~\bibnamefont
  {S\'en\'echal}}, \bibinfo {author} {\bibfnamefont {D.}~\bibnamefont {Perez}},
  \ and\ \bibinfo {author} {\bibfnamefont {M.}~\bibnamefont
  {Pioro-Ladri\`ere}},\ }\href {\doibase 10.1103/PhysRevLett.84.522} {\bibfield
   {journal} {\bibinfo  {journal} {Phys. Rev. Lett.}\ }\textbf {\bibinfo
  {volume} {84}},\ \bibinfo {pages} {522} (\bibinfo {year} {2000})}\BibitemShut
  {NoStop}%
\bibitem [{\citenamefont {B\"uttiker}(1986)}]{PhysRevLett.57.1761}%
  \BibitemOpen
  \bibfield  {author} {\bibinfo {author} {\bibfnamefont {M.}~\bibnamefont
  {B\"uttiker}},\ }\href {\doibase 10.1103/PhysRevLett.57.1761} {\bibfield
  {journal} {\bibinfo  {journal} {Phys. Rev. Lett.}\ }\textbf {\bibinfo
  {volume} {57}},\ \bibinfo {pages} {1761} (\bibinfo {year}
  {1986})}\BibitemShut {NoStop}%
\bibitem [{\citenamefont {Zhu}\ and\ \citenamefont
  {Balatsky}(2003)}]{PhysRevB.67.165326}%
  \BibitemOpen
  \bibfield  {author} {\bibinfo {author} {\bibfnamefont {J.-X.}\ \bibnamefont
  {Zhu}}\ and\ \bibinfo {author} {\bibfnamefont {A.~V.}\ \bibnamefont
  {Balatsky}},\ }\href {\doibase 10.1103/PhysRevB.67.165326} {\bibfield
  {journal} {\bibinfo  {journal} {Phys. Rev. B}\ }\textbf {\bibinfo {volume}
  {67}},\ \bibinfo {pages} {165326} (\bibinfo {year} {2003})}\BibitemShut
  {NoStop}%
\bibitem [{\citenamefont {Knap}\ \emph {et~al.}(2013)\citenamefont {Knap},
  \citenamefont {Arrigoni},\ and\ \citenamefont {von~der
  Linden}}]{PhysRevB.88.054301}%
  \BibitemOpen
  \bibfield  {author} {\bibinfo {author} {\bibfnamefont {M.}~\bibnamefont
  {Knap}}, \bibinfo {author} {\bibfnamefont {E.}~\bibnamefont {Arrigoni}}, \
  and\ \bibinfo {author} {\bibfnamefont {W.}~\bibnamefont {von~der Linden}},\
  }\href {\doibase 10.1103/PhysRevB.88.054301} {\bibfield  {journal} {\bibinfo
  {journal} {Phys. Rev. B}\ }\textbf {\bibinfo {volume} {88}},\ \bibinfo
  {pages} {054301} (\bibinfo {year} {2013})}\BibitemShut {NoStop}%
\bibitem [{\citenamefont {Slater}\ and\ \citenamefont
  {Koster}(1954)}]{PhysRev.94.1498}%
  \BibitemOpen
  \bibfield  {author} {\bibinfo {author} {\bibfnamefont {J.~C.}\ \bibnamefont
  {Slater}}\ and\ \bibinfo {author} {\bibfnamefont {G.~F.}\ \bibnamefont
  {Koster}},\ }\href {\doibase 10.1103/PhysRev.94.1498} {\bibfield  {journal}
  {\bibinfo  {journal} {Phys. Rev.}\ }\textbf {\bibinfo {volume} {94}},\
  \bibinfo {pages} {1498} (\bibinfo {year} {1954})}\BibitemShut {NoStop}%
\end{thebibliography}%

\end{document}